\documentclass[conference]{IEEEtran}
\IEEEoverridecommandlockouts
\usepackage{cite}
\usepackage{amsmath,amssymb,amsfonts}
\usepackage{algorithmic}
\usepackage{graphicx}
\usepackage{textcomp}
\usepackage{xcolor}
\usepackage{booktabs}
\usepackage{lipsum}
\usepackage{multirow}
\usepackage{fixltx2e}
\usepackage{float}
\usepackage{dblfloatfix}
\usepackage{xspace}

\def\BibTeX{{\rm B\kern-.05em{\sc i\kern-.025em b}\kern-.08em
    T\kern-.1667em\lower.7ex\hbox{E}\kern-.125emX}}
\begin{document}

\onecolumn
\begin{flushleft}
“© 2020 IEEE.  Personal use of this material is permitted.  Permission from IEEE must be obtained for all other uses, in any current or future media, including reprinting/republishing this material for advertising or promotional purposes, creating new collective works, for resale or redistribution to servers or lists, or reuse of any copyrighted component of this work in other works.”
\end{flushleft}.
\twocolumn

\title{Representing Gate-Level SET Faults \\ by Multiple SEU Faults at RTL\\

}

\author{\IEEEauthorblockN{Ahmet Cagri Bagbaba\IEEEauthorrefmark{1}\IEEEauthorrefmark{2},
Maksim Jenihhin\IEEEauthorrefmark{2}, Raimund Ubar\IEEEauthorrefmark{2}, Christian Sauer\IEEEauthorrefmark{1}}
\IEEEauthorblockA{\IEEEauthorrefmark{1}Cadence Design Systems,
Munich, Germany; \IEEEauthorrefmark{2} Tallinn University of Technology, Tallinn, Estonia \\
Email: \IEEEauthorrefmark{1}\{abagbaba, sauerc\}@cadence.com,
\IEEEauthorrefmark{2}\{maksim.jenihhin, raimund.ubar\}@taltech.ee}}

\maketitle

\begin{abstract}
The advanced complex electronic systems increasingly demand safer and more secure hardware parts. Correspondingly, fault injection became a major verification milestone for both safety- and security-critical applications. However, fault injection campaigns for gate-level designs suffer from huge execution times. Therefore, designers need to apply early design evaluation techniques to reduce the execution time of fault injection campaigns. In this work, we propose a method to represent gate-level Single-Event Transient (SET) faults by multiple Single-Event Upset (SEU) faults at the Register-Transfer Level. Introduced approach is to identify true and false logic paths for each SET in the flip-flops' fan-in logic cones to obtain more accurate sets of flip-flops for multiple SEUs injections at RTL. Experimental results demonstrate the feasibility of the proposed method to successfully reduce the fault space and also its advantage with respect to state of the art. 
It was shown that the approach is able to reduce the fault space, and therefore the fault-injection effort, by up to tens to hundreds of times.   
\end{abstract}


\section{Introduction}
The fault injection technique is widely used for evaluating functional safety \cite{7753547} and security threats resilience \cite{4302714} in integrated circuits. For safety-critical applications, it is an established, accurate method to assess the effectiveness of the deployed safety mechanisms. For security-critical applications, the technique is efficient to mimic an attack by physical fault injection aimed to alter the program flow or the processed data \cite{8167705}. However, depending on the abstraction level of the circuit and the size of the fault space, a fault injection campaign can be very costly. 

One of the challenges of fault injection campaigns is the vast number of possible fault locations. For a simulation-based fault injection campaign \cite{Ziade03asurvey}, engineers simulate a fault-free design and its copies with faults injected one at a time. This may imply enormous execution times, especially for the gate level fault analysis.  Hence, there is a high demand for methodologies that can support designers in the early-stage design exploration of reliability factors. Moreover, fault injection into gate-level models is quite late in the integrated circuit development cycle, and any design modifications become more expensive in terms of the required engineering effort. Several researchers delved into the early-stage explorations of the designs for both safety and security applications~\cite{6979088, 7035347, 6800420, 6850666}. In both safety and security-related applications, early design evaluation is necessary to minimize design iterations and resources, thus to enable faster design closure times.

In this work, we focus on SET faults at the gate level and propose an efficient solution to represent them by multiple SEU faults at the RT level. The relevance of this problem for safety-critical applications grows with the downscaling of the technology nodes, forcing designers to evaluate system’s safety against SET faults, which affect combinational elements of the circuit. However, this comprehensive evaluation at the gate level is not affordable in terms of the execution time of fault injection campaigns for the industrial-sized designs. From the security point of view, SET faults at the gate level represent laser fault attacks, which can be observed in flip-flops (FFs) as single or multiple errors \cite{6037460}. Here, it is crucial to evaluate laser attacks in order to determine which vulnerable SET faults create single or multiple errors in the sequential elements of the design. 

To tackle the listed problems, we propose a methodology for representing gate-level transient faults, such as SETs, by Multiple Flip-Flop Upset (MFFU) at RTL. In the case of Soft Error Reliability (SER) assessment for safety applications such as automotive, MFFU becomes functionally equivalent for EDA tools to multiple simultaneous SEUs. For vulnerability analysis against fault-injection attacks on security-critical designs, MFFU refers to single and multi-bit fault injections. In this work, first, we identify static fan-in cones of each FF at the gate level. Second, we perform propagation analysis to identify SET faults that have true (sensitizable) paths to FF inputs. In this way, we obtain optimized FF sets as representatives of all SET faults to guide RTL multiple SEU fault injection campaigns. As a result, this method can successfully reduce the fault space and enhance the high complexity of fault injection campaigns. Without loss of generality, the proposed methodology is demonstrated on a Cadence EDA (Electronic Design Automation) tool flow, but it remains applicable to other tool flows as well. The main contribution of this work is as follows:

\begin{itemize}
\item An approach to move the gate-level SET vulnerability analysis to RTL
\item A technique to reduce the fault space at RTL by applying gate-level propagation analysis 
\item A systematic and workload-independent methodology for representing the gate-level SETs by multiple SEUs at RTL supported by industrial-grade EDA tool flow
\end{itemize}

\begin{figure*}[t]
\centerline{\includegraphics[scale=0.85]{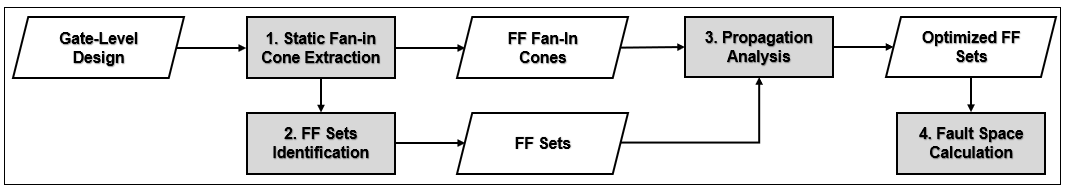}}
\caption{Steps of proposed methodology.}
\label{flow}
\end{figure*}

The rest of the paper is organized as follows. In Section \ref{related}, we give an overview of the related work. The proposed methodology is explained in Section \ref{method}. The experimental results are discussed in Section \ref{results}. Finally, Section \ref{conclusion} concludes the paper.

\section{Related Works}\label{related}

Relevant solutions for the above problem are proposed in \cite{6800420} and \cite{6850666}. However, these state-of-the-art approaches rely on the static cones pre-analysis only and do not consider if a SET fault actually propagates to the FF inputs.~\cite{6800420} proposes an RTL fault injection model which is representative for laser fault attacks. To do that, the authors analyse the circuits structurally and find intersection cones which guide the fault injection in advance. On the other hand, they neither create FF sets that cover all SET faults nor optimize FF sets by considering true/false paths. Similarly,~\cite{6850666} models the locality of a laser attack in case of multiple-bit faults. The authors analyse the circuits structurally as well and, afterwards, create FF sets. However, the authors consider only the supersets and reject all the subsets. In this way, each combination of SEUs in the superset is a trial to hit a fault in any smaller cone intersection. Yet, the probability of hitting a SET in case of any superset by selected random multiple SEU is low.

There are other studies which investigate the impact of SET faults.~\cite{1544534} estimates the impact of SET faults without layout information by identifying a pair of gates in which SET can propagate to multiple outputs.~\cite{1688899} analyzes the impact of SETs through Algebraic Decision Diagrams and Binary Decision Diagrams (BDD) and \cite{5580219} improves this method by considering multiple effects. Finally,~\cite{7373147} suggests performing a stochastic gate-level simulation for small circuits. Last but not least, there are some works that investigate the combination of different fault analysis technologies such as \cite{8949396} and \cite{8854449}. These works combine the strength of formal methods and fault injection simulators; however, they analyse only permanent faults and do not analyse the representation of gate-level SET faults at RTL.

Different from the works listed above, this paper proposes a more efficient technique to prune the fault space by considering the propagation of SET faults. The significant speedup is achieved by running the RTL fault injection procedure on the accurately selected multiple flip-flop upset faults.

\section{Representing gate-level SET Faults by multiple SEU Faults at RTL}\label{method}
In this work, the aim is to identify Multiple Flip-flop Upset sets for RTL fault injection, which represent all gate-level SET faults. By doing so, we reduce the number of injections required to evaluate the effect of SET faults.

\begin{figure}[b]
\centerline{\includegraphics[scale=0.55]{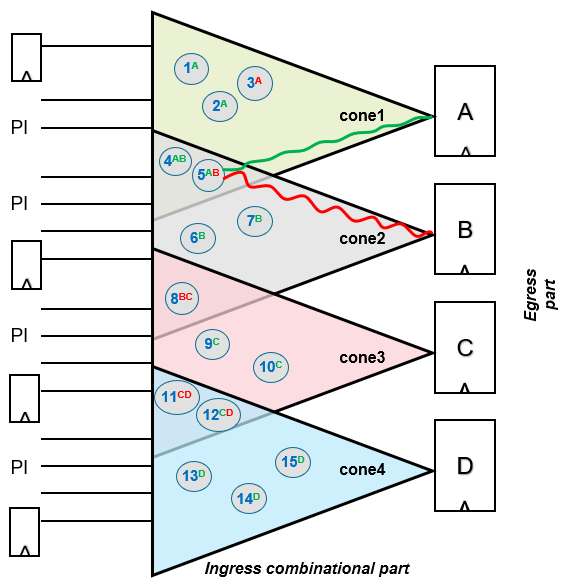}}
\caption{Extracting fan-in cones of each FF and finding propagation paths.}
\label{fanin}
\end{figure}

\begin{table*}[ht]
\centering
\caption{Results of Example Design Given in Fig.~\ref{fanin}}
\label{exampleFFSets}
\begin{tabular}{@{}ccccc@{}}
\toprule
\textbf{Affected Cone} & \textbf{FF Sets} & \textbf{Multiplicity} & \multicolumn{1}{l}{\textbf{Optimized FF Sets}} & \multicolumn{1}{l}{\textbf{Optimized Multiplicity}} \\ \midrule
Cone 1 & A, B    & 2 & A, B & 2 \\
Cone 2 & A, B, C & 3 & A, B & 2 \\
Cone 3 & B, C, D & 3 & C    & 1 \\
Cone 4 & C, D    & 2 & D    & 1 \\ \bottomrule
\end{tabular}
\end{table*}

The SET fault model implies flipping the value of a signal in the combinational cloud and holding the value for a specified period of time. SEU fault model implies flipping the value of the output of a sequential element and holding it until it is overwritten with new data. SEUs can be applied on the outputs of sequential elements, such as memories, FFs and latches. We apply SET faults for one clock cycle length. The proposed flow is shown in Fig.~\ref{flow} and starts with the (1) extraction of static fan-in cones of each FF in gate-level netlist. In the next step (2), FF sets are created to represent each SET faults on the fan-in cones of FFs. Then, we perform propagation analysis (3) to check if SET faults propagate to the FF inputs. If a SET fault does not propagate, then we check if this changes created FF sets. In this way, we obtain optimized FF sets, which are representative of all SET faults, which propagate to the FF inputs. Finally (4), we calculate the fault space to see the reduction when compared to state-of-the-art and random multi-bit injection approaches. The following subsections explain each step of the proposed method in detail.

\subsection{Static Fan-in Cone Extraction of Flip-Flops at gate level}
As a first step, we extract fan-in cones of each FF at the gate level, as it is illustrated in Fig.~\ref{fanin}. In the beginning, we generate a list of all faults in the design. Then, we extract fan-in information from all FFs in the ingress combinational part of the design. Each fan-in cone search starts from a FF and expands backward, i.e. in the direction of inputs of the combinational cloud until it encounters a FF output or a primary input (PI). Finally, all SETs in each cone are enumerated to map each SET to a FF set. This step is performed by using Cadence{\textregistered} JasperGold Functional Safety Verification App. 
\subsection{Flip-Flop Sets Identification}
The second step of the proposed methodology is the identification of FF sets, which will be used as a MFFU injection target in the following steps. To do that, we consider each fan-in cone independently and determine FF sets, which cover all possible scenarios, as shown in the second column of Table~\ref{exampleFFSets}. For instance, if cone-1 is affected by a SET fault, we can cover this SET fault by injecting multiple MFFUs on A and B because cone-1 has an intersection with cone-2 which is the fan-in cone of B. This process is repeated for each cone, and FF sets are obtained with a size between 1 (in case the cone does not intersect with any other cones) and N FFs (in case all cones have an intersection). 

Extracted FF sets are flip-flops of the circuit potentially affected by a SET. Therefore, MFFU injection can be limited to this set of FF. Table~\ref{exampleFFSets} also shows the multiplicity information of each FF set. The multiplicity of a FF set is the number of FF in a set. For instance, if a SET fault occurs in cone-2, it can propagate to the A, B, C FFs, causing different combinations of upsets on this set. This means that the less is the number of FF in a set (less multiplicity), the higher is the probability of hitting a real MFFU. We will use this information in the following steps. Moreover, multiplicity is important for the calculation of fault space, which will be given in the next sections. It is obvious that there are 8 combinations in one FF set with a multiplicity 3.

\begin{figure}[b]
\centerline{\includegraphics[scale=0.35]{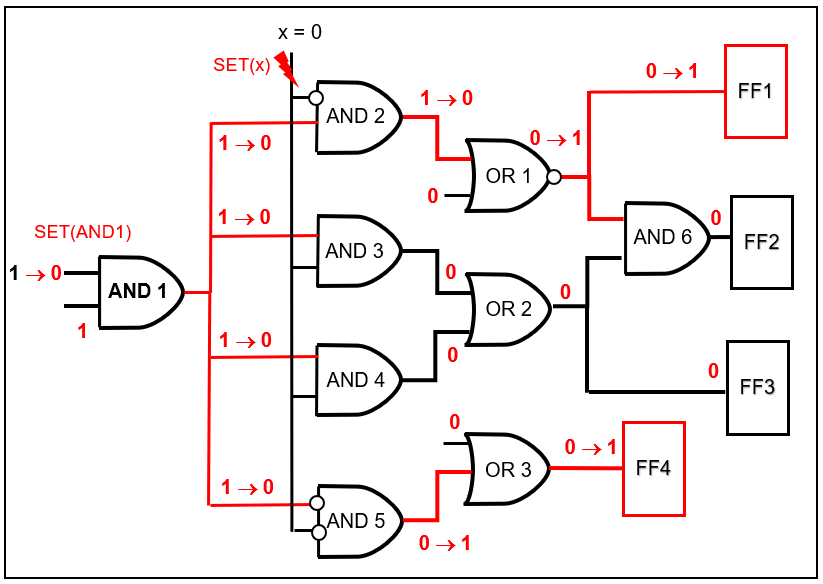}}
\caption{Motivational example to find propagated and not-propagated faults.}
\label{lowLevelEx}
\end{figure}

\subsection{Propagation Analysis}
In this work, unlike state-of-the-art researches, we also take propagation of faults into consideration in order to reduce fault space more. For this step, we deploy the formal techniques to investigate the behaviour of a design under fault. The theory behind formal techniques is creating of Boolean function representation of a design under test so that formal proves can be used. In order to achieve better performance in the modern formal tools, BDDs~\cite{10.1007/11757283_4} and Multiway Decision Graphs (MDGs)~\cite{Corella96multiwaydecision} are widely used. 

The formal analysis deploys formal methods to determine the propagation of faults. Propagation analysis verifies if there is a combination of inputs that provoke fault propagation. If a fault propagates to FF inputs, we accept that the fault has a true path to FF inputs. Otherwise, it has a false path and should be excluded from the analysis. In this step, formal properties to perform the analysis are automatically generated and verified with respect to all possible input stimuli.

The simple and high-level example in Fig.~\ref{fanin} illustrates that there are some SET faults in the intersection cones with a false path to the FF inputs. In this figure, green paths and superscripts point the true paths (fault propagates) while red ones show that the related fault has a false path (fault does not propagate). As a result of this step, we obtain optimized FF sets, as shown in the fourth column of Table~\ref{exampleFFSets}. It is obvious that some larger FF sets are disappeared due to non-observable faults that cannot be propagated. In this way, optimized multiplicities are obtained along with the reduced number of FF sets in some circuits. This step is performed by using Cadence{\textregistered} JasperGold Functional Safety Verification App. In the following subsection, we show a more detailed motivational example for the propagation analysis.

\emph{Motivational Example: Removing the paths which cannot be propagated}\label{propagationExplanation}

To explain the propagation analysis in detail, we use a motivational example given in Fig.~\ref{lowLevelEx} which has fan-out nodes. The circuit includes an input \textbf{x}, and outputs of the gates \textbf{AND1}, \textbf{OR1} and \textbf{OR2}. The SETs may be simulated only for these fan-outs. The steps of the approach can be listed as follows: 
\begin{itemize}
\item Static fan-in cone analysis gives us the following FF sets of MFFU faults: (1, 2, 3, 4) for \textbf{x}, (1, 2, 3, 4) for \textbf{AND1}, (1, 2) for \textbf{OR1}, (2, 3) for \textbf{OR2}.
\item After removing of duplicated sets, we get the initial sets of MFFU faults: (1, 2, 3, 4), (1, 2), (2, 3). 
\item By propagation analysis, we see that for SET on AND1 we never reach all FFs, rather only either (1, 4) or (2, 3) due to the fact that the propagation of a SET at \textbf{AND1} is controlled by signal \textbf{x}=0 (by blocking two of four AND gates). Therefore, the superset (1, 2, 3, 4) for \textbf{AND1} should be replaced by subsets (1, 4) and (2, 3). In other saying, SET(\textbf{AND1}) is mapped to (1, 4) and (2, 3) FF sets.
\item Moreover, the SET on the input \textbf{x} is always blocked either on \textbf{AND5} (if output of \textbf{AND1}=1), or on \textbf{AND2}, \textbf{AND3}, \textbf{AND4} (if output of \textbf{AND1}=0). Hence, the superset (1, 2, 3, 4) for SET(x) should be replaced by (1, 2, 3).
\item As a result, we get instead of initial {(1, 2, 3, 4), (1, 2), (2, 3)}, optimized FF sets {(1, 2), (2, 3), (1, 4) (2, 3), (1, 2, 3)}, where (2, 3) can be removed as it is duplicated. 
\item Thus, the final optimized FF sets: (1, 2), (2, 3), (1, 4), (1, 2, 3).
\end{itemize}

In this motivational example, we analyzed the propagation of SETs only on \textbf{x} and the outputs of \textbf{AND1}, \textbf{OR1} and \textbf{OR2}. 
The propagation analysis is sufficient for the SETs at these four locations that also represent the remaining SET faults in the fan-out free regions.

\subsection{Fault Space Calculation}

In the fault injection procedure, SEUs are injected in all possible locations and at each clock cycle~\cite{8906932}. Therefore, the number of injections required for a single transient fault is large, especially for the industrial-sized designs. When considering the size and low speed of fault injection simulations at the gate level, optimization methods should  be applied. Hence, considering the huge number of SET injections at the gate level, our proposed method significantly reduces the number of injections by identifying optimized FF sets when compared to state-of-the-art and random multi-bit injection approaches applied in safety and security applications.

Our proposed methodology can significantly reduce the fault space by leveraging the FF sets with propagation analysis. In this work, we compare our results with the state-of-the-art and random multi-bit injection. State-of-the-art researches such as~\cite{6800420} and \cite{6850666} rely on only a static approach and do not consider the propagation analysis. Similarly, the random multi-bit injection method considers all possible FF combinations. In order to calculate fault space or the number of injections, we use the following equation where N is the number of FFs, k$_1$, k$_2$, ..., k$_N$ are the numbers of FF in each set and 1$\leq$k$_i$$\leq$N. given in~\cite{6850666}.
\vspace{-1mm}
\begin{equation}\label{formula1}
{FaultSpace}_{Total}=\sum_{i=1}^{N}{(2^{k_i}-1)}
\end{equation}

By using the above equation, the total fault space for the example given in Fig.~\ref{lowLevelEx} can be calculated effortlessly. As it is explained in Section~\ref{propagationExplanation}, we have the initial and not-optimized sets which represent the state-of-the-art approach as (1, 2, 3, 4), (1, 2) and (2, 3). By using the given formula, the total number of faults is 21. On the other hand, we have optimized FF sets as (1, 2), (2, 3), (1, 4), (1, 2, 3), which require 16 number of injections. Therefore, our proposed method can reduce the total fault space from 21 to 16 for the motivational example given in Fig.~\ref{lowLevelEx}.

\begin{table*}[hbt!]
\centering
\caption{Experimental Results: Fault Spaces Achieved by Three Methods}
\label{ExpResults}
\resizebox{\textwidth}{!}{%
\begin{tabular}{@{}cc|cccc|cccc|c@{}}
\toprule
\multirow{2}{*}{\textbf{Circuit}} &
  \multirow{2}{*}{\textbf{\# FF}} &
  \multicolumn{4}{c|}{\textbf{without propagation analysis}} &
  \multicolumn{4}{c|}{\textbf{with propagation analysis}} &
  \textbf{Random Multi-Bit Injection} \\ \cmidrule(l){3-11} 
 &
   &
  \textbf{\# sets} &
  \textbf{\# superset} &
  \textbf{max multiplicity} &
  \textbf{Total Faults} &
  \textbf{\# sets} &
  \textbf{\# superset} &
  \textbf{max multiplicity} &
  \textbf{Total Faults} &
  \textbf{Total Faults} \\ \cmidrule(r){1-2}
\textbf{b01}         & 5   & 5  & 2  & 4  & 1.80E+01 & 3  & 1  & 4  & 1.50E+01    & 3.20E+01 - 1 \\
\textbf{b02}         & 4   & 4  & 1  & 3  & 7.00E+00 & 1  & 1  & 2  & 3.00E+00    & 1.60E+01 - 1 \\
\textbf{b03}         & 30  & 8  & 3  & 12 & 4.14E+03 & 3  & 1  & 9  & 5.11E+02    & 1.07E+09 - 1 \\
\textbf{b04}         & 66  & 27 & 10 & 19 & 4.00E+06 & 5  & 4  & 8  & 1.02E+03    & 7.38E+19 - 1 \\
\textbf{b05}         & 34  & 62 & 2  & 33 & 9.00E+09 & 61 & 5  & 31 & 2.00E+09    & 1.72E+10 - 1 \\
\textbf{b06}         & 8   & 7  & 5  & 4  & 4.30E+01 & 7  & 5  & 4  & 4.30E+01    & 2.56E+02 - 1 \\
\textbf{b07}         & 46  & 51 & 2  & 35 & 4.00E+10 & 43 & 3  & 26 & 8.00E+07    & 7.04E+13 - 1 \\
\textbf{b08}         & 21  & 19 & 2  & 18 & 2.70E+05 & 11 & 2  & 18 & 2.62E+05    & 2.10E+06 - 1 \\
\textbf{b09}         & 28  & 14 & 1  & 28 & 3.00E+08 & 7  & 1  & 27 & 1.00E+08    & 2.68E+08 - 1 \\
\textbf{b10}         & 17  & 45 & 9  & 11 & 5.91E+03 & 13 & 4  & 11 & 2.62E+03    & 1.31E+05 - 1 \\
\textbf{b11}         & 31  & 43 & 9  & 18 & 4.65E+05 & 9  & 2  & 16 & 6.60E+04    & 2.15E+09 - 1 \\
\textbf{b13}         & 50  & 40 & 13 & 13 & 9.15E+03 & 20 & 9  & 9  & 9.47E+02    & 1.13E+15 - 1 \\ \bottomrule
\end{tabular}%
}
\end{table*}

\section{Experimental Results and Discussion}\label{results}
In order to verify the effectiveness of proposed methodology, we evaluate our methodology on the ITC'99~\cite{867894} benchmark circuits.

In order to perform fan-in cone analysis and propagation analysis, we deploy Cadence tools along with the developed script sets, which execute on gate-level design. Meanwhile, all applied methods remain applicable to other tool flows. In the beginning, we synthesize Verilog or VHDL design through Cadence{\textregistered} Genus{\texttrademark} Synthesis Solution to obtain gate-level representation of the design. Then, steps 1, 2 and 3 shown in Fig.~\ref{flow} are performed on our application which deploys Cadence{\textregistered} JasperGold Functional Safety Verification App.

We use three methods to show the fault space reduction and compare the results. The first method is "without propagation analysis" which represents the state-of-the-art as in~\cite{6850666}. The main difference between our proposed methodology "with propagation analysis" and the state-of-the-art is the identification of true (sensitizable) paths. We leverage the analysis by identifying SET faults which do not propagate to FF inputs so that fault space is reduced more. In other words, we cut down the pessimism in the results. The third approach used for comparison is "Random Multi-Bit injection". This is basically injecting faults on all possible combinations of FFs randomly that naturally causes huge fault space. Our application is capable of building the fault space for each method and given design without any significant effort.

\begin{figure*}[hbt!]
\centerline{\includegraphics[scale=0.9]{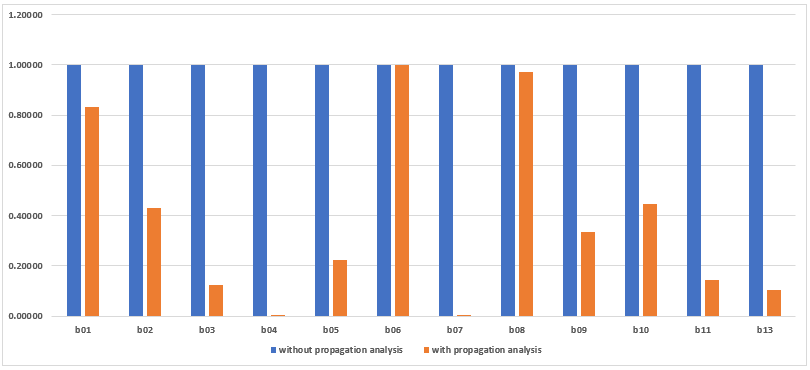}}
\caption{Fault Space comparison.}
\label{graph1}
\end{figure*}

All experimental results are presented in Table~\ref{ExpResults}. The selected designs include various designs from the ITC'99 benchmark. During creating of FF sets, we remove faults on clock and reset signals from the analysis due to the fact that the clock tree is not known in this stage of the design. Other faults except clock and reset are kept as they are. This step is done in our application automatically. We show the number of sets, number of supersets, maximum multiplicity and calculated fault spaces for each analysis and design. The number of sets shows the number of all identified FF sets before duplicated ones are removed. In contrast, the number of supersets points the same after duplicated ones are removed. Total Faults are calculated by using the Equation~\ref{formula1}. 

In Table~\ref{ExpResults}, it can be seen that our proposed methodology reduces the Total Faults significantly when compared to both state-of-the-art and the random multi-bit injection approaches. For some circuits such as \textbf{b01} and \textbf{b08}, we are able to reduce only the number of supersets while the maximum multiplicity is still the same in both cases. Moreover, there is no optimization achieved in \textbf{b06}. For the rest of the circuits given in Table~\ref{ExpResults}, we both optimize the number of supersets and maximum multiplicity. Thereby, the total set of faults  are optimized significantly, as shown in Fig.~\ref{graph1} (values are normalized). It is observable that total faults in the proposed methodology (orange bars) are less than the other two methods. We also add that we reduce the fault space from 1.20 times to a few hundred times when compared without propagation analysis, depending on the circuit.

Moreover, we also compare our results with the well-known Statistical Fault Injection (SFI) approach~\cite{5090716} in case initial population sizes calculated before are used. SFI can be used for transient fault injection campaigns to reduce the execution times while keeping a meaningful number of injections with an error margin. This is one of the possible ways to perform RTL fault injection campaigns after FF sets are defined by using the methodology presented in this paper. In an SFI campaign, the sample size or the margin of the error with a certain confidence level are determined by using the Equation~\ref{SFIformula} defined in~\cite{5090716}. In this way, it is possible to obtain precise results while injecting a small number of faults~\cite{5090716}. The technique allows to know the margin of error while restricting the campaign time to the minimum. To sum up, there are three confidence levels in SFI as 90\%, 95\%, and 99.8\%. In this work, we only use the 95\% confidence level as it is the one that is practically used in the industry. Also, three error margins are defined as 5\%, 1\% and 0.1\%. 

\begin{equation}\label{SFIformula}
n=\frac{N}{1+e^{2} \times (\frac{N-1}{t^{2} \times p \times (1-p)}) }
\end{equation}

In Table~\ref{SFI_Results}, we show the SFI results. In this table, N shows the initial population. In our case, N is equal to the total faults shown in Table~\ref{ExpResults}. Moreover, n(5\%), n(1\%) and n(0.1\%) show the required sample size with the error margins 5\%, 1\% and 0.1\% respectively. This shows that our proposed methodology can prune the fault space from 1.12 times to a few hundred times in case faults are injected by using SFI. Note, the results for some sample sizes remain similar due to the fact that the initial population is always finite. Even so, we show that a significant reduction is achieved by using the proposed methodology, especially when we reduce the error margins. Therefore, it is efficient to use the proposed methodology and to select a sample for fault injection among the pre-defined initial populations in the MFFU space identified using the method "with propagation analysis".

\begin{table*}[]
\centering
\caption{Comparison of three methods in a SFI Campaign with 95\% Confidence Level}
\label{SFI_Results}
\resizebox{\textwidth}{!}{%
\begin{tabular}{@{}ccccc|cccc|cccc@{}}
\toprule
\multirow{2}{*}{\textbf{Circuit}} &
  \multicolumn{4}{c|}{\textbf{without propagation analysis}} &
  \multicolumn{4}{c|}{\textbf{with propagation analysis}} &
  \multicolumn{4}{c|}{\textbf{Random Multi-Bit Injection}} \\ \cmidrule(l){2-13} 
 &
  \textbf{N} &
  \textbf{n(5\%)} &
  \textbf{n(1\%)} &
  \textbf{n(0.1\%)} &
  \textbf{N} &
  \textbf{n(5\%)} &
  \textbf{n(1\%)} &
  \textbf{n(0.1\%)} &
  \textbf{N} &
  \textbf{n(5\%)} &
  \textbf{n(1\%)} &
  \textbf{n(0.1\%)} \\ \cmidrule(r){1-1}
\textbf{b01} & 1.80E+01 & 1.70E+01 & 1.80E+01 & 1.80E+01 & 1.50E+01 & 1.40E+01 & 1.50E+01 & 1.50E+01 & 3.20E+01 - 1 & 3.00E+01 & 3.20E+01 & 3.20E+01 \\
\textbf{b02} & 7.00E+00 & 7.00E+00 & 7.00E+00 & 7.00E+00 & 3.00E+00 & 3.00E+00 & 3.00E+00 & 3.00E+00 & 1.60E+01 - 1 & 1.50E+01 & 1.60E+01 & 1.60E+01 \\
\textbf{b03} & 4.14E+03 & 3.52E+02 & 2.89E+03 & 4.12E+03 & 5.11E+02 & 2.20E+02 & 4.85E+02 & 5.11E+02 & 1.07E+09 - 1 & 3.84E+02 & 9.60E+03 & 9.60E+05 \\
\textbf{b04} & 4.00E+06 & 3.84E+02 & 9.58E+03 & 7.74E+05 & 1.02E+03 & 2.79E+02 & 9.22E+02 & 1.02E+03 & 7.38E+19 - 1 & 3.84E+02 & 9.60E+03 & 9.60E+05 \\
\textbf{b05} & 9.00E+09 & 3.84E+02 & 9.60E+03 & 9.60E+05 & 2.00E+09 & 3.84E+02 & 9.60E+03 & 9.60E+05 & 3.44E+10 - 1 & 3.84E+02 & 9.60E+03 & 9.60E+05 \\
\textbf{b06} & 4.30E+01 & 3.90E+01 & 4.30E+01 & 4.30E+01 & 4.30E+01 & 3.90E+01 & 4.30E+01 & 4.30E+01 & 2.56E+02 - 1 & 1.54E+02 & 2.49E+02 & 2.56E+02 \\
\textbf{b07} & 4.00E+10 & 3.84E+02 & 9.60E+03 & 9.60E+05 & 8.00E+07 & 3.84E+02 & 9.60E+03 & 9.49E+05 & 7.04E+13 - 1 & 3.84E+02 & 9.60E+03 & 9.60E+05 \\
\textbf{b08} & 2.70E+05 & 3.84E+02 & 9.28E+03 & 2.11E+05 & 2.62E+05 & 3.84E+02 & 9.27E+03 & 2.06E+05 & 2.10E+06 - 1 & 3.84E+02 & 9.56E+03 & 6.59E+05 \\
\textbf{b09} & 3.00E+08 & 3.84E+02 & 9.60E+03 & 9.57E+05 & 1.00E+08 & 3.84E+02 & 9.60E+03 & 9.51E+05 & 2.68E+08 - 1 & 3.84E+02 & 9.60E+03 & 9.57E+05 \\
\textbf{b10} & 5.91E+03 & 3.61E+02 & 3.66E+03 & 5.88E+03 & 2.62E+03 & 3.35E+02 & 2.06E+03 & 2.62E+03 & 1.31E+05 - 1 & 3.83E+02 & 8.95E+03 & 1.15E+05 \\
\textbf{b11} & 4.65E+05 & 3.84E+02 & 9.41E+03 & 3.14E+05 & 6.60E+04 & 3.82E+02 & 8.39E+03 & 6.18E+04 & 2.15E+09 - 1 & 3.84E+02 & 9.60E+03 & 9.60E+05 \\
\textbf{b13} & 9.15E+03 & 3.69E+02 & 4.69E+03 & 9.06E+03 & 9.47E+02 & 2.74E+02 & 8.62E+02 & 9.46E+02 & 1.13E+15 - 1 & 3.84E+02 & 9.60E+03 & 9.60E+05 \\ \bottomrule
\end{tabular}%
}
\end{table*}

\section{Conclusions}\label{conclusion}
In this work, we propose a methodology to represent gate-level SET faults by multiple SEU faults at RTL. It enables a solution for the high complexity problem of expensive gate-level fault injection campaigns by changing the abstraction level. We improve the state-of-the-art by considering propagation analysis of each SET fault. First, we find static fan-in cones of each FF at the gate level. Second, FF sets are created pessimistically, meaning that propagation analysis is not considered. Third, we execute propagation analysis by using a formal approach to find SET faults that propagate to FF inputs. Then, optimized FF sets are created again with less pessimism. Finally, we calculate the fault space to show the effectiveness of the proposed methodology. In this way, we significantly reduce the number of fault injections and obtain a higher probability of hitting a true multiple SEU fault. Experimental results show that we make the fault space smaller by up to tens to hundreds of times. 

As future work, we aim to apply this methodology for functional safety and security evaluation in industrial-sized CPU designs. 

\section*{Acknowledgment}
This research was supported by project RESCUE funded from the European Union's Horizon 2020 research and innovation programme under the Marie Sklodowaska-Curie grant agreement No 722325.

\bibliographystyle{IEEEtran}
\bibliography{main}

\end{document}